# ChordMics: Acoustic Signal Purification with Distributed Microphones

Weiguo Wang, Jinming Li, Meng Jin, Yuan He
School of Software and BNRist, Tsinghua University
{wwg18, li-jm19}@mails.tsinghua.edu.cn, mengj@mail.tsinghua.edu.cn, heyuan@tsinghua.edu.cn

***Abstract*—Acoustic signal acts as an essential input to many systems. However, the pure acoustic signal is very difficult to extract, especially in noisy environments. Existing beamforming systems are able to extract the signal transmitted from certain directions. However, since microphones are centrally deployed, these systems have limited coverage and low spatial resolution. We overcome the above limitations and present ChordMics, a distributed beamforming system. By leveraging the spatial diversity of the distributed microphones, ChordMics is able to extract the acoustic signal from arbitrary points. To realize such a system, we further address the fundamental challenge in distributed beamforming: aligning the signals captured by distributed and unsynchronized microphones. We implement ChordMics and evaluate its performance under both LOS and NLOS scenarios. The evaluation results tell that ChordMics can deliver higher SINR than the centralized microphone array. The average performance gain is up to 15dB.**

## I. Introduction

Acoustic signal acts as an essential input to many systems, such as speech recognition, security monitoring, malfunction detection, etc. A nature requirement underlying all these systems is that the collected acoustic signal should be pure and strong enough to support high-quality information extraction on the upper layer. This prohibits the application of these systems in noisy environments. However, many acoustic signal based applications are exactly required to work in noisy environments. A typical example is malfunction detection, which usually works in industrial scenarios [1], [2].

In malfunction detection systems, the characteristics of the sound generated by a machine can be exploited as an indicator of malfunction. However, most factories have multiple machines distributed in a large workshop, which operate and roar simultaneously and loudly. In this case, the acoustic signal of the target machine is seriously interfered by those of the other machines, which undermines the characteristics of the target signal. To get pure signal, a straightforward method is to deploy an acoustic sensor exactly next to each target machine. However, without purification, the collected signal is still a mixture of the sound from different machines. As an example, we in [3] provide the acoustic signal collected by the sensor deployed close to a certain machine in a power plant. In this example, we can find that the signal is quite mussy and mixed. It is difficult to extract the signal from any source.

In this paper, we ask - can we design a signal purification system which can extract the acoustic signal from an exact source, as shown by Fig. 1(a). Beamforming is a promising technique to achieve this vision. This method utilizes a microphone array for directional signal reception. Specifically, it combines the signals received by different microphones in a way that signals from a particular direction experience constructive interference while the others experience destructive interference. In this way, it achieves signal purification.

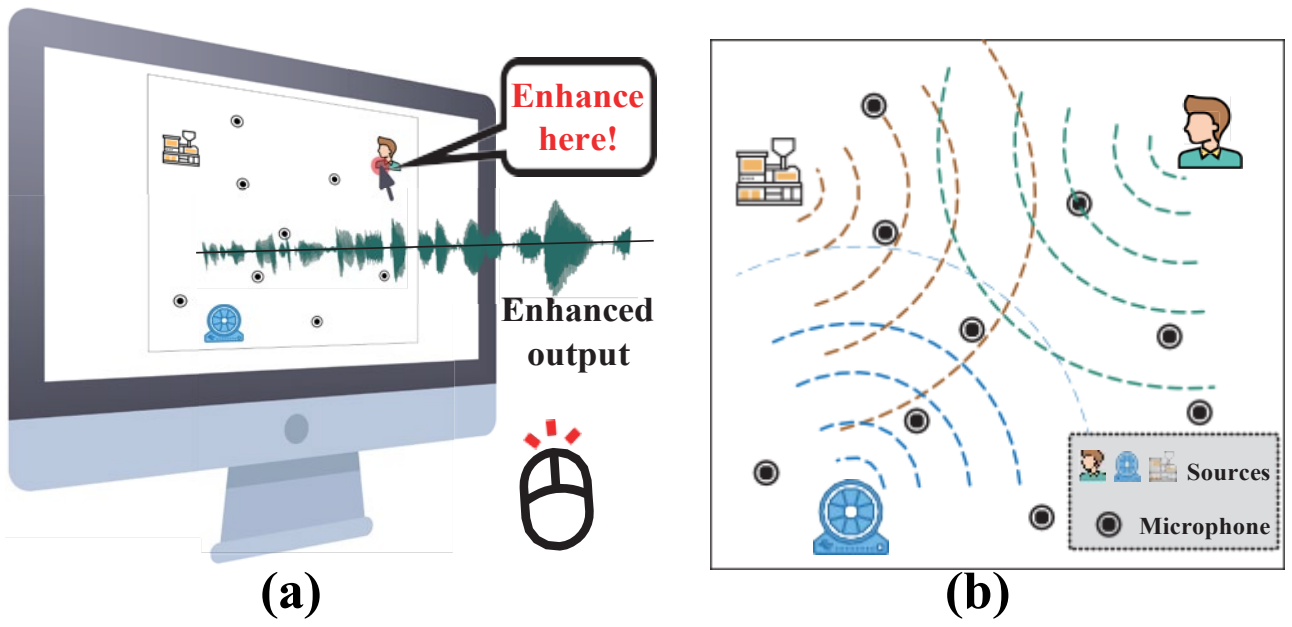

**Fig. 1: (a)** Enhance any point of the sound field by simply clicking the screen. **(b)** The deployment of ChordMics.

However, traditional beamforming systems have the following two inherent limitations:

- **Low spatial resolution.** Traditional beamforming systems form a directional beampattern. They cannot separate sources whose directions of arrival (DoA) are close. Therefore, they cannot accurately control the extracted target.
- **Limited coverage.** Due to the centralized deployment, traditional beamforming systems provide limited coverage and are vulnerable to obstacles. So, they may leave many blind spots, especially in large-scale and complex environments like stores and factories.

We in this paper present ChordMics, a system that uses *distributed* beamforming to achieve highly controllable signal extraction in multipath-rich scenarios. Fig. 1(b) illustrates the layout of ChordMics. Different from the centralized microphone array, the distributed microphones can exploit the spatial diversity of the microphones to achieve full coverage of the environment. Additionally, ChordMics enables flexible scalability: we can scale up or down the coverage on demand by simply adding or removing microphones. Most importantly, ChordMics achieves controllable signal enhancement. Specifically, in ChordMics, signals form the same source arrive at microphones from different directions. By intersecting signals from different directions, we can form a beampattern which could point to a certain point rather than a direction, thus yielding higher spatial resolution.

However, underlying the distributed beamforming is a fundamental challenge: how to accurately estimate the *relative delays* of the desired sources among different microphone pairs? The estimation can be easily achieved in centralized beamforming where the relative positions of the microphones are accurately determined, and the clock is shared. In a distributed scenario, however, accurate microphone coordinates are not available due to inevitable measurement errors during deployment. Moreover, tight time synchronization is hard to achieve among distributed nodes where microphones are deployed [4], and the non-negligible time offset, as well as the unstable sound speed, would result in additional estimation error.

ChordMics solves this problem based on a blind alignment method, which estimates signal's relative delay by directly mining the "timestamp" information carried by the signal itself. Specifically, although received by the microphones on different locations, the signals from the same source exhibit identical signal pattern. So, by simply computing the correlation between signals received by different microphones, we can get the signal's relative delay: the time-domain location of the correlation peak gives an accurate estimation of the relative delay. Yet, although this method achieves high resolution, it brings ambiguity: multiple acoustic sources will produce multiple correlation peaks. It is difficult to find the peak (and the corresponding relative delay) that corresponds to the desired signal. Fortunately, we find that i) the error in the distance-based delay estimation is bounded; ii) compared with RF signal, the propagation speed of acoustic signal is quite slow. So, two closely located sources can still produce very different signal delays, making the correlation peaks sparsely locate on the time domain. Therefore, by quantifying the error boundary and using it as a filter on the correlation result, we can find out the desired peak and resolve the ambiguity.

The contributions of this paper are summarized as follows:

- We propose ChordMics, a system that realizes distributed beamforming for controllable acoustic signal purification. ChordMics contributes a solution to the core challenge of many acoustic-signal-based approaches, ranging from sensing to communication techniques.
- To achieve distributed beamforming in practice, we present a novel blind signal alignment method, which achieves perfect signal alignment based on the "timestamp" information carried by the signal itself, without relying on tight synchronization among microphones or the accurate geometry of the distributed microphones.
- We implement and evaluate ChordMics with extensive experiments, which demonstrate the high coverage and spatial resolution of ChordMics.

Our paper is organized as follows. Sec. II introduces the background, challenges and overview of ChordMics. Sec. III presents the design details. Sec. IV discusses some practical issues. We evaluate the performance in Sec. V. Sec. VI summarizes the related work, and Sec. VII draws the conclusion.

## II. Background and Challenges

### A. Delay-and-Sum Beamformer

Beamforming is a signal processing technique used in receiver arrays (i.e., microphone arrays in our case) for directional signal reception. The enhancement is achieved by combining the signals received by different microphones in a way that signals from particular direction experience constructive interference while others experience destructive interference [5]. The key to beamforming is to estimate the *relative arriving delays* of the desired signals on different microphone pairs, based on which we can align and enhance the desired source.

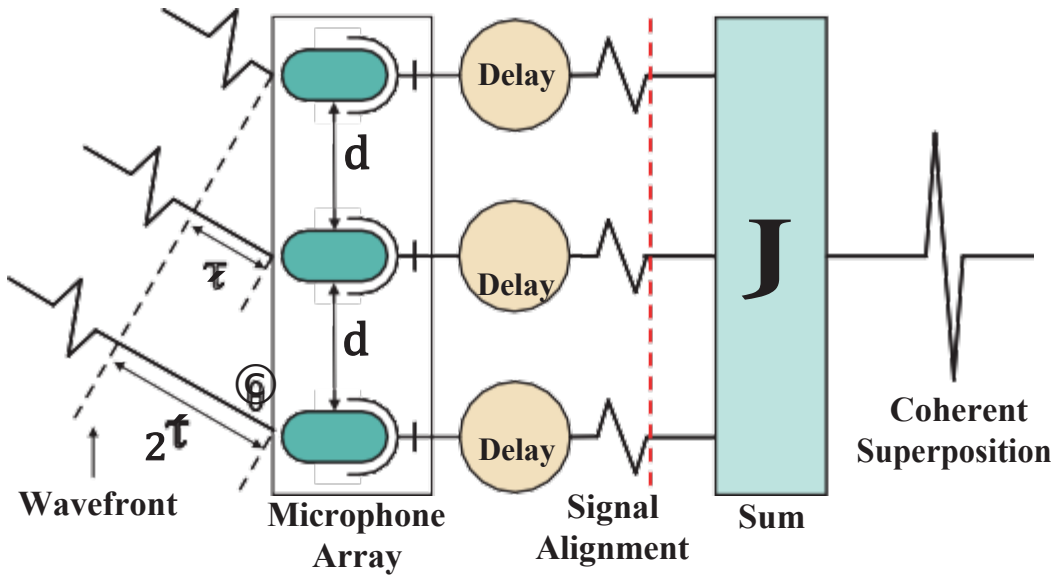

**Fig. 2:** Delay-and-Sum Beamformer.

Fig. 2 shows a typical beamforming technique, named Delay-and-Sum Beamformer (DSB) [6]. In DSB, the relative delay is estimated based on the known geometry of the microphones. Specifically, if the desired source is far away from the microphone array, we can treat the propagating rays of signals as parallel. Meanwhile, The microphones are spaced equally with inter-distance $d$. Then, as shown in Fig. 2, we can estimate the relative delay (denoted by $\tau$) between two adjacent microphones as $\tau = \frac{d\cos(\theta)}{c}$, where $c$ denotes the speed of sound and $\theta$ denotes the DoA of the signals.

By compensating the arriving delay of the signals from a specific direction, we could accurately align those signals and thus the signals can be constructively combined. On the contrary, the interfering signals from other directions remain misaligned, and thus will be canceled.

### B. Distributed Beamforming Challenges

Although DSB systems can achieve signal purification, it suffers limitations like low resolution and limited coverage. By contrast, distributed beamforming can solve this problem leveraging the spatial diversity of the microphones and the controllable beampattern.

However, accurate estimation of the relative delays is not trivial in distributed beamforming. Recall that in DSB systems, the relative delay is estimated based on the known special geometry of the centralized microphone array. In distributed beamforming, however, the microphones are deployed separately. To estimate relative delays, we require both tight synchronization among microphones and the accurate geometry of the microphones. Specifically, for a signal from a certain source $S_D$, its relative arriving delay on two microphones $M_A$ and $M_B$ can be estimated as

$$\tau^D_{AB} = \frac{\Delta d^D_{AB}}{c} = \frac{d^D_A - d^D_B}{c} \tag{1}$$

where $d^D_A$ and $d^D_B$ denote the distance between $M_A$ and $S_D$, and $M_B$ and $S_D$. $\Delta d^D_A = d^D_A - d^D_E$ denotes the difference between distances from the desired source $S_D$ to the two microphones $M_A$ and $M_B$. In practice, however, estimating relative delays based on the above equation is inaccurate due to the following three reasons: First, accurate alignment requires a millimeter-level distance accuracy, which is not available due to inevitable measurement errors. Second, the speed of the sound $c$ is not stable and will change with the temperature, which introduces additional errors. More importantly, signals are recorded by different devices with inevitable clock offsets. This leads to inconsistent timestamps of signals, which directly offsets the estimation result. In summary, the above distance-based method provides only a coarse estimation of the relative delays, which cannot support accurate signal alignment.

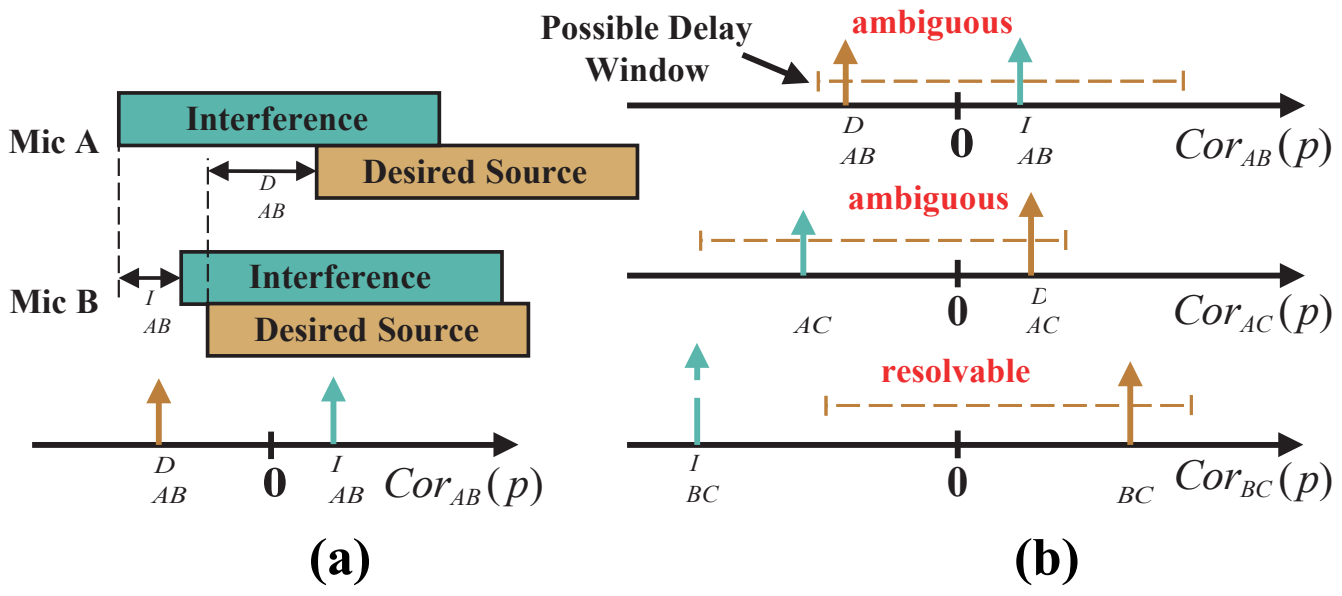

**Fig. 3: (a)** Illustration of cross-correlation function **(b)** Possible Delay Window can be used as a filter

An alternative method is to use the cross-correlation Function(CCF) between signals. Fig. 3(a) plots an example where two microphones (denoted as $M_A$ and $M_B$) record signals from two sources ($S_D$ and $S_I$ ). The bars with different colors represent the signals from different sources. We correlate the signals captured by two microphones with different offsets $p$. The correlation result is near zero, except when the signals from the same sources are perfectly aligned. As shown in Fig. 3(a), we can find two peaks in the correlation result. The locations of the two peaks on time domain indicate the relative delays of the corresponding signal sources.

However, although the CCF-based method provides fine-grained measurements of the relative delays, it also brings ambiguity: there are multiple peaks and we cannot tell which peak corresponds to the relative delay of the desired source $S_D$. Such a problem becomes more challenging in the multipath-rich case, where multiple copies of the the same signal arrive at a microphone with different delays. In this case, just one source can produce multiple correlation peaks, making it more difficult to find out the desired peaks.

### C. ChordMics Overview

Based on the above analysis, we can summarize that the key challenge we face in estimating the relative delays is to resolve the ambiguity while maintaining the high resolution. ChordMics is able to solve this conundrum. The design of ChordMics is based on an observation that although the error of distance-based method is intolerable for alignment, it can be bounded. By quantifying the boundary and applying it as a filter to the CCF result, we can find the desired peaks and remove ambiguity. Thus, our design of ChordMics adopts a multi-resolution approach, which has two main components.

In the first component, we *bound* the error introduced by the distance-based method and *quantify* the boundary of this error. Specifically, we first leverage a reference signal to compensate clock offsets among different microphones, which are originally boundless and are the main error of the distance-based method. Then, based on a detailed analysis of how different factors affect the estimation of the delay, we quantify the boundary of the error. The boundary provides a potential range in which the desired peak would locate in. We term this range as Possible Delay Window (PDW).

The second component uses the PDW as a filter to find out the desired peaks. We use the case in Fig. 3(b) to explain this method. In this case, we have three microphones ($M_A$, $M_B$, and $M_C$) and two sources ($S_D$ and $S_I$ ). By computing the correlation between signals recorded by each pair of microphones (i.e., $\langle M_A, M_B\rangle$, $\langle M_A, M_C\rangle$ and $\langle M_B, M_C\rangle$), we can get three sets of correlation peaks. For each microphone pair, if the distance between the desired peak and the interference peaks is large enough (e.g., $\tau^I_B$ and $\tau^D_B$ in Fig. 3(b)), we can successfully find out the desired peak $\tau^D_B$ since only one peak locates in the PDW.

However, for some microphone pairs (e.g.$\langle M_A, M_B\rangle$ and $\langle M_A, M_C\rangle$), we will find multiple peaks in one PDW. This occurs when the two signal sources share similar relative distance to the two microphones. In ChordMics, we solve this problem leveraging the spatial diversity of the microphones and the relationship between signal's relative delay on different microphone pairs. Specifically, we find that for any three microphones ($M_A$, $M_B$ and $M_C$), there is an identical relation among the relative delays of the desired source $S_D$ to three microphone pairs, that is $\tau^D_B = \tau^D_{AC} - \tau^D_{AB}$. Based on this relation, once the desired peak corresponding to just one microphone pair is identified (e.g.,$\langle M_B, M_C\rangle$), all the other desired peaks (and thus the corresponding relative delays) can be identified iteratively. In fact, due to the spatial diversity of the distributed microphones, we empirically find that the possibility of finding such an "anchor" microphone pair (e.g., $\langle M_B, M_C\rangle$) reaches 98.3% when inter-distances between the sources are larger than 0.5m.

## III. Design

### A. Stage 1: Coarse-Grained Alignment

The first task of ChordMics is to narrow the error range of the distance-based method and then quantify the error boundary. Recall that the error is composed of three parts: inaccurate distance measurement, uncertain sound speed and time offset among microphones. Among them, time offset is the dominant factor. Without a tight synchronization, the time offset continuously increases due to the different stepping rates of the clocks, which leads to boundless error. Therefore, to narrow the error bound, we should first remove the time offset.

To solve this problem, we make use of a certain reference acoustic source to eliminate the uncertain time offset among

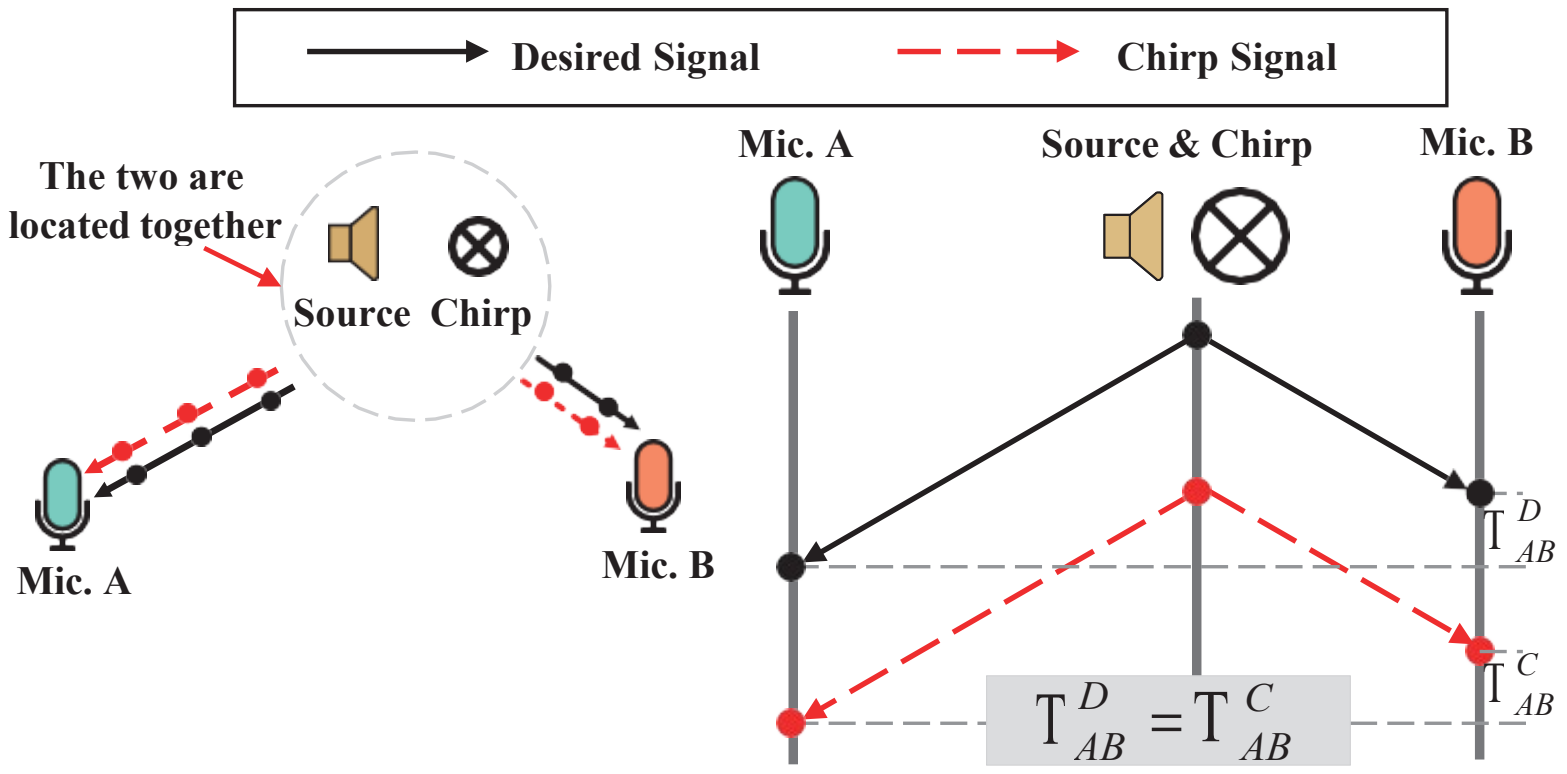

**(a)** Layout **(b)** Timeline

**Fig. 4:** A simple case where the chirp device and the desired source are located together.

microphones. To better understand this idea, let us first look at a simple example.

*1) A simple case:* In this case, we attach the desired source with an additional device which periodically broadcasts signals with a known pattern (e.g., an acoustic chirp signal in our design), as shown in Fig. 4(a). This additional signal acts as a reference to detect the relative arriving delay of the signals from the desired source to the microphones. Specifically, since the pattern of the reference signal is known, we can detect the relative delay of the reference signal directly based on *template matching*. Then, since the external source and the desired source are in the same location, their signals exhibit exactly the same arriving delays to the microphones, like the example shown in Fig. 4(b). So, we can accordingly obtain the relative delays of the desired signals and achieve signal alignment. Note that by referring to the reference signal, we can directly align the desired signals, which avoids the use of Eq. (1) for delay estimation and thus naturally eliminates the uncertainties mentioned in Sec.II-B.

In detail, we will take the following two steps to align and enhance the desired signals:

First, suppose we have $N$ microphones $\mathbf{M} = \{M_1, ..., M_N\}$, and the signal received by each microphone $M_n$ ($1 \leq n \leq N$) is denoted as $x_n$. For each microphone $M_n$, we use a signal template to detect the time domain location of the chirp symbol in $x_n$. The detected location is denoted as $t_n$. Now, without loss of generality, if we treat the signal's arriving time on microphone $M_1$ as a reference point, the relative delay of the chirp signals (and also the desired signal) on each microphone pair $\langle M_1, M_n \rangle$ is given by $t_n - t_1$.

Second, we shift each signal series $x_n$ with the corresponding relative delay $t_n - t_1$. Then the enhanced signal can be calculated by adding all $N$ signals aligned by the chirp as:

$$y(t) = \sum_{n=1} x_n(t - (t_n - t_1))$$

.

Here, we want to explain why we use the chirp symbol as the reference signal. The main reason is its high sensitivity to signal misalignment. Specifically, the chirp signal is a sinusoidal signal whose frequency varies linearly with time over the available bandwidth. Therefore, even a slight misalignment of two chirp signals can lead to a significant decrease in the auto-correlation result, resulting in a particularly narrow correlation peak and thus highly accurate alignment.

*2) General Cases:* In the simple case, an additional chirp device is required to be attached to each desired source. This requirement is however unrealistic in practice because the acoustic source which the user interests in can be everything in the environment, like a conversation between two people, or even the sound of the wind. One cannot expect all the targets are equipped with a additional signal source.

Now, we consider a more realistic case where the desired sources and the chirp device are deployed separately. In this case, the relative delay of the desired signal is no longer equal to that of the chirp signal. The offset between them is caused by the difference in their propagation distance to the microphones. Therefore, to further get the relative delay of the desired signal, we should compensate for the above mentioned difference in propagation distance.

In the following of this section, we use an example with two microphones (denoted by $M_A$ and $M_B$) to illustrate this idea in detail, as shown in Fig. 5(a). We denote the desired source and the chirp device by $S_D$ and $S_C$ and the relative delays of their signals on the two microphones by $\tau_A^D$ and $\tau_A^C$ (as shown in Fig 5(b)). Recall that $\tau_A{}^C$ can be easily obtained based on chirp detection. To derive $\tau_A^D$ , we only need to estimate and compensate for the difference between $\tau_A^D$ and $\tau_A^C$ . We denote such difference as $\Delta_{AB}$.

By comparing Fig. 5 and Fig. 4, we find that the main cause of the difference between $\tau_A{}^D$ and $\tau_A{}^C$ is the signals' different propagation distance to the two microphones. Specifically, based on Eq.(1), we know that

$$\begin{aligned} \tau^C_{AB} &= \frac{\Delta d^C_{AB}}{c} + \delta_{AB} = \frac{d^C_A - d^C_B}{c} + \delta_{AB} \\ \tau^D_{AB} &= \frac{\Delta d^D_{AB}}{c} + \delta_{AB} = \frac{d^D_A - d^D_B}{c} + \delta_{AB} \end{aligned} \quad (3)$$

where $\Delta d_A^C$ and $\Delta d_A^D$ respectively denote the difference between the distance from $S_C$ and $S_D$ to the two microphones. $\delta_{AB}$ is the clock offset between $M_A$ and $M_B$. Therefore, by

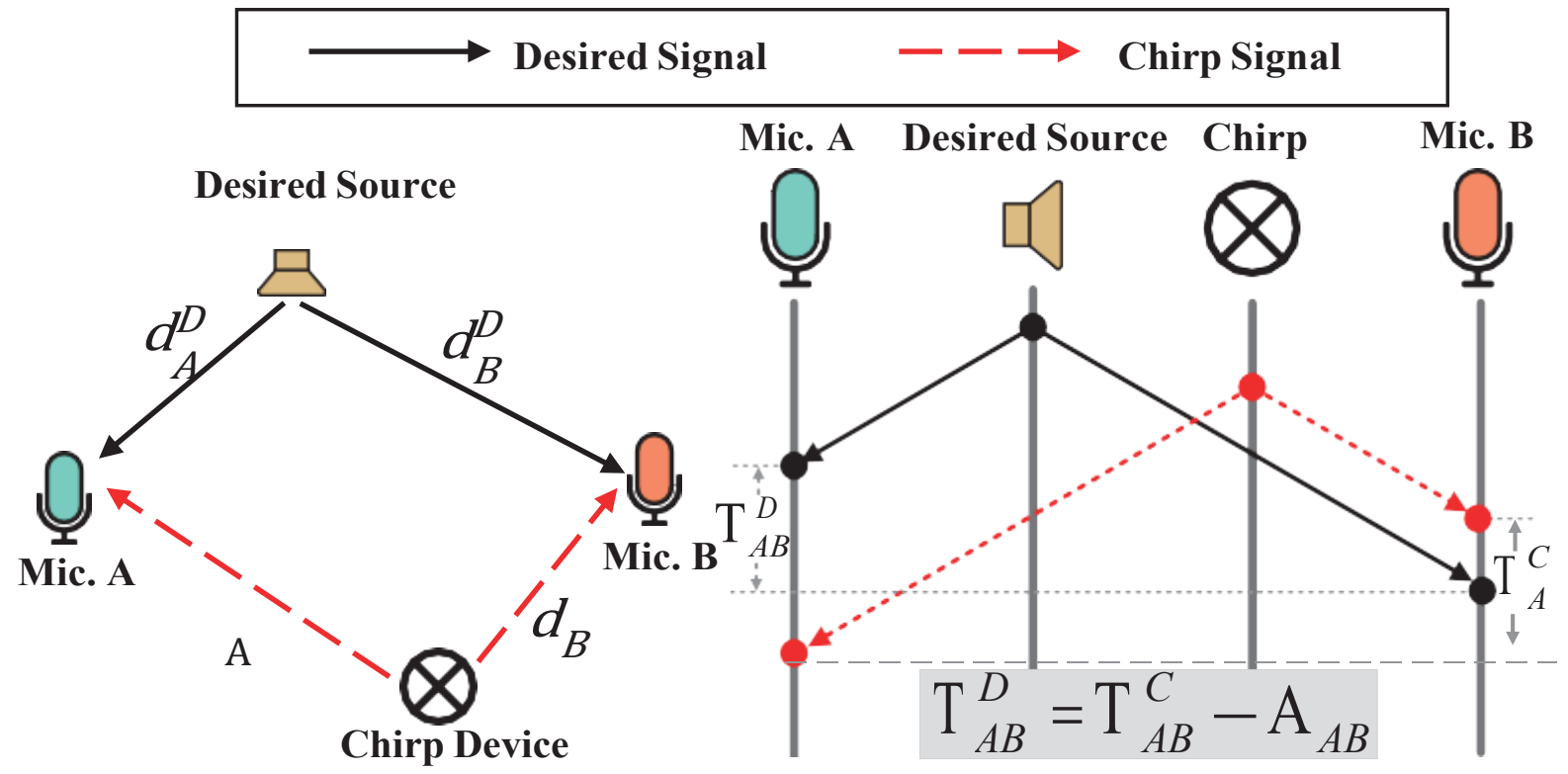

**(a)** Layout **(b)** Timeline

**Fig. 5:** A general case where the chirp device and the desired source are not in the same place

subtracting Eq.(3) from Eq.(2), we can remove the clock offset $\delta_{AB}$ and get $\Delta_{AB}$:

$$\Delta_{AB} = \frac{\Delta d^D_{AB} - \Delta d^C_{AB}}{c} = \frac{d^D_A - d^D_B - d^C_A + d^C_B}{c} \quad (4)$$

Then we can estimate $\tau^D_A$ as:

$$\tau^D_{AB} = \tau^C_{AB} + \frac{d^D_A - d^D_B - d^C_A + d^C_B}{c} \quad (5)$$

We now try to quantify the error boundary of $\tau^D_A$. Recall that $\tau^C_A$ can be accurately estimated based on chirp detection. So, to quantify the error boundary of $\tau^D_A$, we only need to consider the estimation error of $\Delta_{AB}$. Clearly, after removing the clock offset, there are only two factors that can cause the estimation error: inaccurate distance measurement and the uncertain sound speed. We denote the distance error as $e_d$ and the maximum and minimum of sound speed as $c_{min}$ and $c_{max}$. Then the error of $\Delta_{AB}$ (denoted by $e_{\Delta_{AB}}$) is bounded by:

$$e_{\Delta_{AB}} \leq \frac{4e_d}{c_{min}} + \left| d^C_A - d^D_A - d^C_B + d^D_B \right| \left( \frac{1}{c_{min}} - \frac{1}{c_{max}} \right) \quad (6)$$

where the first term represents the error introduced by distance measurement, and the second item represents error introduced by the uncertain sound speed.

In summary, Eq. (6) tells that in the general cases, we can still eliminate the unknown clock offset between the microphones by referring to the chirp signal. While to compensate $\Delta_{AB}$ (which is related to the propagation distance and speed of the signal), we introduce errors caused by inaccurate distances measurement and uncertain sound speed. Fortunately, however, these two types of errors are limited and estimable.

We use a concrete example to show the estimation of $e_{\Delta_{AB}}$ in practice. Assume the system is deployed in a 20m×20m room and the distance measurement error $e_d$ is 0.2m (which is exaggerated in reality). Thus we have $|d^C_A - d^D_A - d^C_E + d^D_E| \leq 20\sqrt{2}$ m. Since the sound speed typically ranges from 337m/s to 348m/s, we can further estimate based on Eq. (6) that $e_{\Delta_{AB}} \leq$5ms.

The above analysis shows that by simply using the inaccurate coordinates and rough sound speed (340m/s), we can bound the error $e_{\Delta_{AB}}$ below 5ms. Such an error range, although still insufficient for signal alignment (where 0.5ms accuracy is required to align 1KHz signals), it can serve as a coarse filter for the following fine-grained alignment.

### *B. Stage 2: Fine-Grained Alignment*

In the previous section, we only get a coarse-grained estimation (i.e., a range) of the signal's relative delays on different microphone pairs, which cannot support accurate signal alignment. In this section, we introduce fine-grained signal alignment based on cross-correlation.

Again, we consider an example with two microphones $M_A$ and $M_B$. The cross-correlation function [7] (CCF) between the received signal $x_A$ and $x_B$ is defined as

$$Cor_{AB}(p) = E\left[x_A(t + p)x_B(t)\right] \quad (7)$$

where $p$ is the displacement between the two signals. For a source $S_D$, when $p = \tau^D_A$, the signals from $S_D$ will be perfectly aligned, and $Cor_{AB}(p)$ will produce a peak on $p = \tau^D_A$.

Ideally, we can directly obtain $\tau^D_A$ based on the time-domain location of the peak as $\tau^D_A = \arg\max_p Cor_{AB}(p)$. In practice, however, there might be multiple signal sources in one environment, which produce different peaks on the correlation result. We can not identify the peak corresponding to the desired source. To understand this problem, let's consider a case with two sources (the desired source $S_D(t)$ and the interference source $S_I(t)$). The signal received at the microphones $M_n \in \{M_A, M_B\}$ is:

$$x_n(t) = \alpha^D_n S_D(t - \tau^D_n) + \alpha^I_n S_I(t - \tau^I_n) + v_n(t)$$

where $\alpha^D_n$ and $\alpha^I_n$ are the attenuation factors of the desired source and the interference, $\tau^D_n$ and $\tau^I_n$ are delays from the two sources to the microphone $M_n$, and $v_n(t)$ denotes the noise. Then, the CCF between $x_A(t)$ and $x_B(t)$ is given by

$$Cor_{AB}(p) = \alpha^D_A \alpha^D_B Cor_{S_D}(p - \tau^D_{AB}) + \alpha^I_A \alpha^I_B Cor_{S_I}(p - \tau^I_{AB}) \quad (8)$$

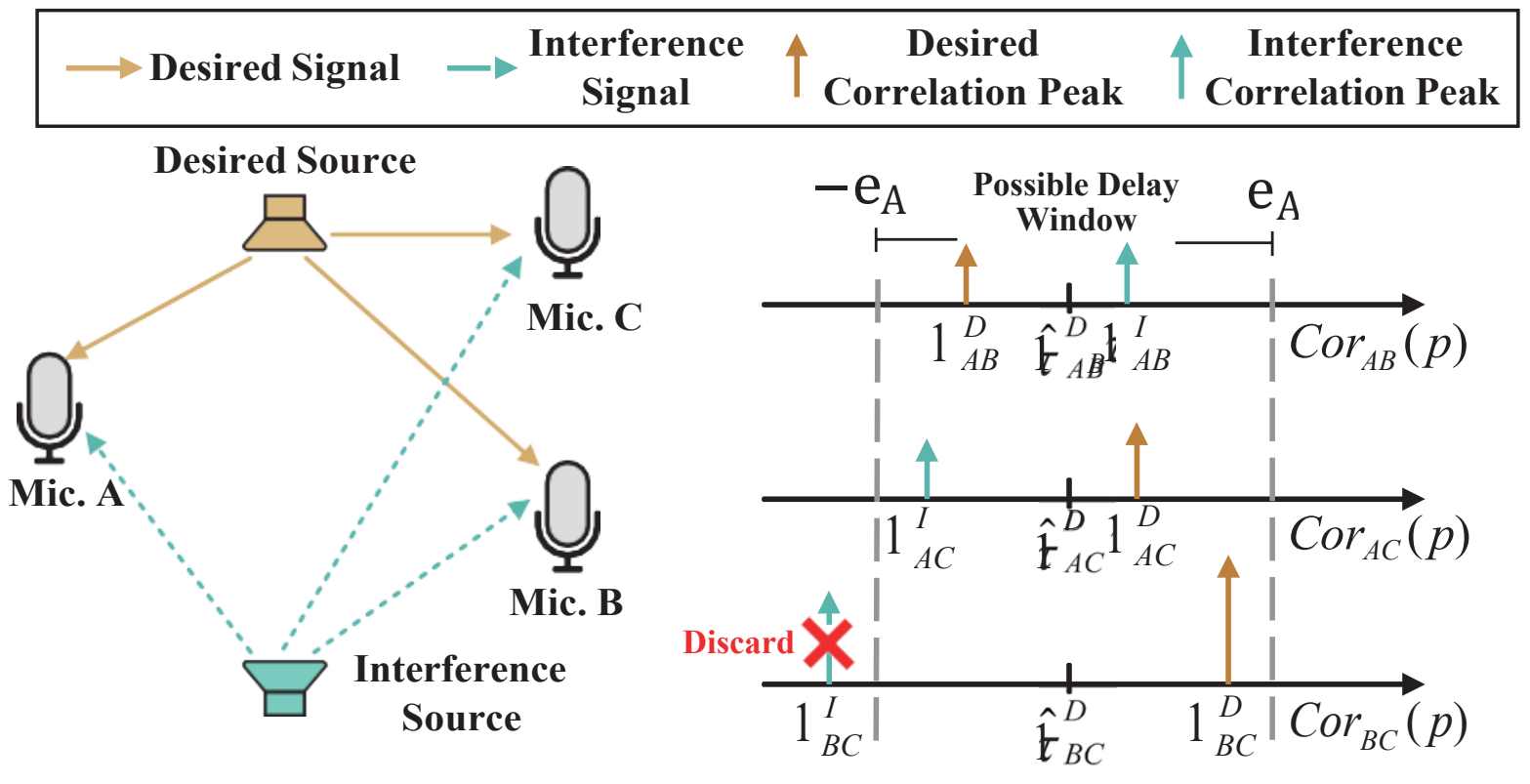

**(a)** Layout **(b)** CCFs with PDW

**Fig. 6:** ChordMics uses PDW and the relation of relative delays to identify the desired peak

where $Cors_{S_D}$ and $Cors_{S_I}$ are self-correlation functions of the desired source $S_D$ and the interference $S_I$, $\tau^D_A = \tau^D_A - \tau^D_E$ and $\tau^I_A = \tau^I_A - \tau^I_E$ are the relative delays of the desired source and the interference(The sources are mutually independent and the noise is uncorrelated with sources and other noise). Eq. (8) tells that $Cor_{AB}$ does have two peaks located at $\tau^D_A$ and $\tau^I_A$ .

So, how to resolve such ambiguity? Fortunately, we have shown in Sec. III-A that it is possible to provide a coarse range of the $\tau^D_A$ , which can be used as a filter on the correlation result to remove the ambiguity. We term such a range as **Possible Delay Window** (PDW), which is defined as $[\tau^D_{AB} - e_{\Delta_{AB}}, \tau^D_{AB} + e_{\Delta_{AB}}]$. $\tau^D_{AB}$ is estimated as

$$\tau^D_{AB} = t_B - t_A + \frac{\Delta_{AB}}{c}$$

where $t_A$ (or $t_B$) is the timestamp of chirp arriving at $M_A$ (or $M_B$), which is stamped by the clock of $M_A$ (or $M_B$).

We use PDW to filter out undesired peaks. Specifically, if there is only one peak within the PDW, this peak must be the desired peak. However, in some rare cases, there might be multiple peaks within one PDW. To resolve this problem, we propose a *successive disambiguation* method that identifies peaks iteratively by exploiting the geometric diversity of microphones. Specifically, consider the case shown in Fig. 6(a). Suppose we have three microphones $M_A$, $M_B$, and $M_C$, and the desired signal arrives at the three microphones at time $t^D_A$, $t^D_B$ and $t^D$. Then, the relative delay of the desired source between the pair $\langle M_A, M_B\rangle$, can be derived as

$$\tau^D_{AB} = t^D_A - t^D_B = (t^D_A - t^D_C) - (t^D_B - t^D_C) = \tau^D_{AC} - \tau^D_{BC}. \quad (9)$$

Eq. (9) provides an identical relation between the relative delays of the desired source $S_D$ among three microphone pairs. Based on this relation, once the desired peak corresponding to just one microphone pair is identified (e.g., $\tau^D_B$ in Fig. 6(a)), we can find another two desired peaks (e.g., $\tau^D_A$ and $\tau^D_A$ ) whose locations satisfy the above relation. In a more general case with more than three microphones, we can iteratively identify all the desired peaks (and thus the corresponding relative delays) on each microphone pair, as long as there is one microphone pair whose PDW contains only one peak.

The process of successive disambiguation can be summarized as follows: i) Calculating the cross-correlation between signals received by each microphone pair; ii) Finding out the microphone pair(s) whose PDW contains only one peak; iii) Identifying all the other desired peaks iteratively based on the identified peak(s) and the relation in Eq. (9).

**Analysis.** One might wonder how likely there exists one microphone pair whose PDW contains only one candidate peak? We find that it occurs as long as the distance between the desired source and each interference source is larger than 0.5m. Specifically, for one microphone pair, e.g. $\langle M_A, M_B\rangle$, once the distance between the desired source and the interference source is less than

$$SR = c \cdot e_{\Delta_{AB}} \quad (10)$$

their peaks will fall in one PDW. We call such a distance as the spatial resolution (SR) of ChordMics. Once the distance between a pair of sources below SR, we might not identify the desired peaks.

We conduct groups of simulations to evaluate how likely ChordMics can separate the desired signal from the interference. In the simulation, the deployment of the microphones is shown in Fig. 7(a). The sources are randomly deployed in the environment with two different densities, which is achieved by controlling the minimum distance between sources at 0.5m and 1.5m. Under each density, we traverse all the possible deployment of the sources and run 12,250,000 sets of experiments. The results show that the possibility that we can separate the desired signal is 98.3% or 99.7% when the minimum distances between the sources are 0.5m or 1.5m, respectively.

Note that even in the extreme case where we cannot identify the desired peak in any cross-correlation results, we can still classify the peaks into groups based on Eq. (9). Peaks in the same group belong to the same acoustic source. The only problem here is we cannot map each group to the corresponding source. ChordMics solves this problem by directly enhancing all these sources and provides them to the user. The user can further manually identify the desired source.

## IV. Practical Issues

### A. *Weighted Combination and Microphone selection*

In distributed beamforming, microphones are deployed separately and the signal strength of a certain source will be quite different on different microphones. According to maximal ratio combining theory [8], [9], to maximize the output SNR, we have to weight the received signals in each microphone in proportion to the signal strength of the desired source. However, we cannot get the signal strength accurately because besides the desired source, the signal is also mixed with the interference and noise. We notice that high cross-correlation value actually indicates high signal strengths, so we estimate combination weights in proportion to the value of the desired peaks in cross-correlation result.

Meanwhile, some microphones might be far away from the desired source. Their signals provide only little contribution in signal enhancement and may incurs higher interference if they locate near the interference source. In this case, we directly discard the signal of these microphones.

### B. *Multiple Sources*

We use two sources to introduce ChordMics above. Actually, ChordMics can be easily extended to the scenarios with more sources. The only difference is that we may have more than two peaks in the correlation result of each microphone array. We can still use the method introduced in Sec. III to resolve such ambiguity and find the desired peak.

## V. Evaluation

### A. *Experimental Setup*

The hardware of ChordMics includes three major parts: wireless node, microphone sensors and PC server. We use the Raspberry Pi 3 Model B+ with on-board WiFi as the wireless nodes. We connected each microphone to a Raspberry Pi through the USB sound-card acting as a wireless microphone (12 in total), and connect a buzzer to a Raspberry Pi acting as a chirp device. The sample rate $F_s$ of microphone is set to 44.1KHz. Both the desired source and the interference are played from JBL PS3300 Speakers. The volume is set to 50 out of 100 (around 70dB) to avoid non-linear distortion. The audio signals collected from microphones are streamed to a MacBook Pro through WiFi interface. All the proceedings like signal detection, alignment, and enhancement are performed on the laptop. We compare ChordMics against two centralized microphone arrays:

- **Commercial-off-the-Shelf (COTS) array.** ReSpeaker 6-Mic Circular Array running DSB and Minimum Variance Distortionless Response (MVDR) beamforming [10];
- **Self-Made 12-Mic array.** Since COTS arrays rarely have more than 6-7 microphones, we use a uniform linear array with 12 microphones running DSB for further comparison.

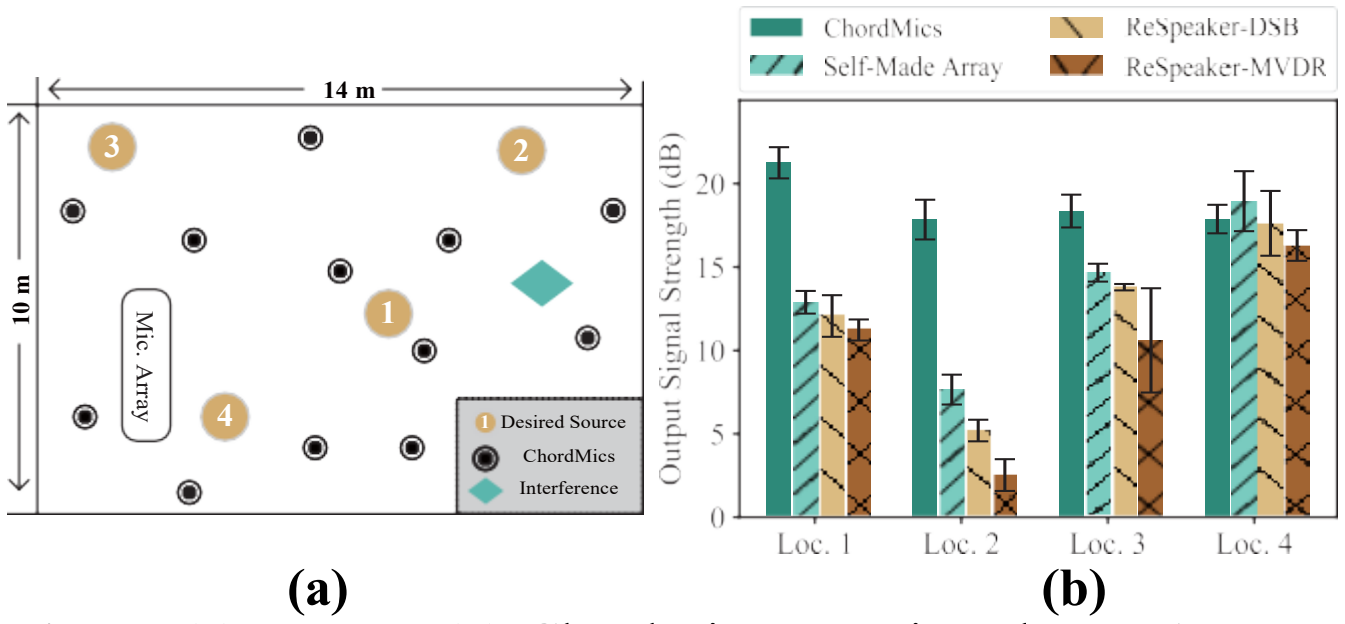

**Fig. 7: (a)** Layout. **(b)** ChordMics vs. Microphones Arrays.

### B. *Output SINR*

To evaluate the enhancement performance of ChordMics, we first observe the SINR of the output signal. In the experiment, we randomly deploy 12 microphones in a room with the area of 10×14 $m^2$. The background noise level is about 42-45dB. The desired sources locate in four different locations of the room, as shown in Fig. 7(a). The rectangle marks the location of centralized arrays. Fig. 7(b) shows the SINRs of four desired sources enhanced by ChordMics and three baseline methods.

We find that ChordMics achieves stable performance across all the four locations of the desired source. In contract, the microphone array achieves satisfying performance only when the desired source locates very near to the array (e.g., on Loc 4). ChordMics outperforms three baselines by 5.3 dB, 6.7 dB and 8.7 dB, respectively. This owning to the distributed deployment characteristic of ChordMics, which provides more opportunities to enhance the signal transmitted from different locations using spatial diversity of the microphones. We believe that it's possible to further improve the performance by combining centralized microphone arrays with ChordMics. Namely, instead of single microphone, each distributed device of ChordMics can be equipped with a centralized microphone array. We leave it to our future work.

### C. *Additional SNR Gain*

As we have discussed in the previous subsection, one advantage of ChordMics is it can leverage the spatial diversity of multiple microphones for robust signal enhancement. In this experiment, we observe how the number of microphones affects the performance of ChordMics. Fig. 8 shows the additional SNR gain (with respective to the single microphone case) under different numbers of microphones. We evaluate the performance in enhancing two types of real-world sources: human voice and engine sound. The energy of the voice and the engine mainly distributes over the frequency range of 50Hz∼ 3.2KHz and 20Hz  0.8KHz, respectively. As expect, the SNR gain increases as the number of microphones increases. There is a slight performance decrease when we add the 9-th microphone. This is because the 9-th microphone is quite close to the interference, which introduces high level of interference. This problem can be solved using the microphone selection method introduced in Sec. IV-A.

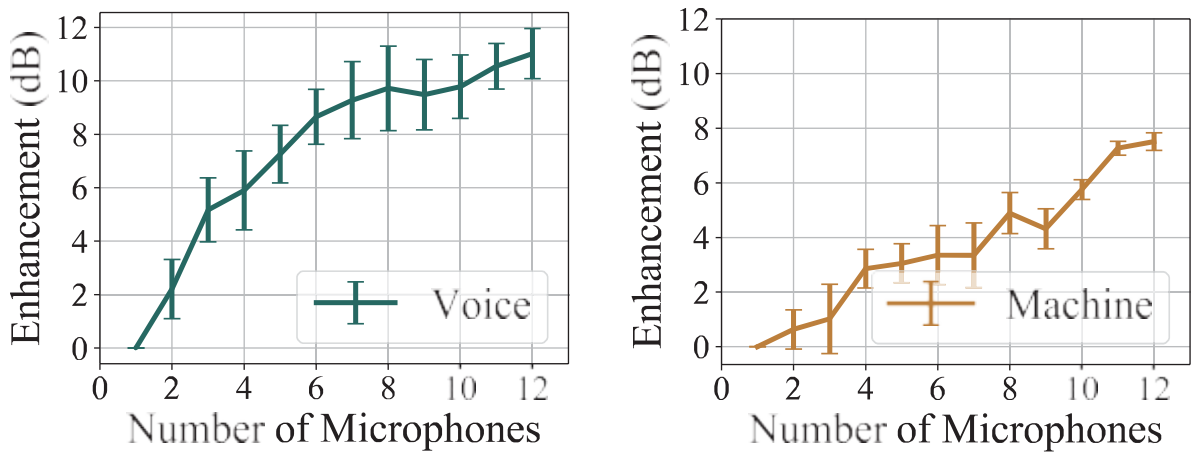

**Fig. 8:** Enhancement Performance of ChordMics

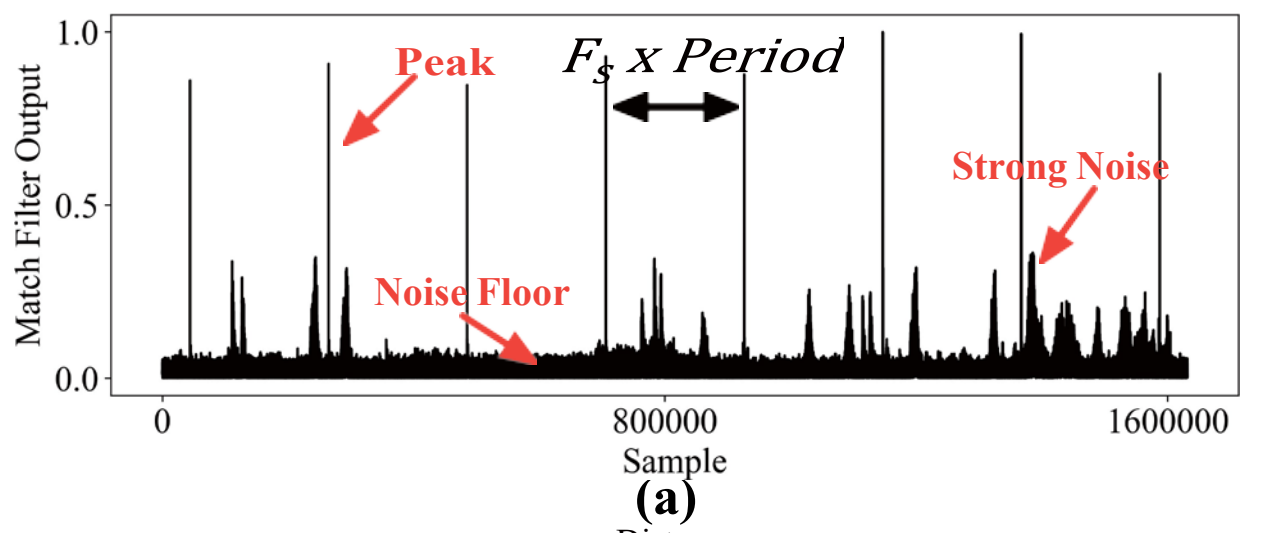

Distance

| Chirp Duration | 3m | 6m | 9m | 12m | 15m | 18m | 21m | 24m | 27m | 30m | 33m | 36m | 39m | 42m | 45m |
|---|---|---|---|---|---|---|---|---|---|---|---|---|---|---|---|
| 1s | 81.7 | 71.8 | 69.0 | 67.7 | 68.0 | 60.0 | 50.6 | 46.8 | 44.7 | 50.9 | 46.9 | 47.4 | 49.7 | 35.4 | 43.5 |
| 0.5s | 77.3 | 66.1 | 63.3 | 62.6 | 63.2 | 56.6 | 47.1 | 43.6 | 53.5 | 52.6 | 41.7 | 39.8 | 47.4 | 38.5 | 38.2 |
| 0.1s | 65.7 | 62.3 | 58.1 | 54.3 | 57.6 | 54.1 | 36.6 | 35.6 | 44.2 | 48.3 | 39.9 | 31.5 | 41.6 | E | 33.7 |
| 0.05s | 69.1 | 58.2 | 52.8 | 42.1 | 49.9 | 50.6 | 36.4 | 31.6 | 36.0 | 41.6 | 37.3 | 36.0 | 42.4 | 35.0 | 33.5 |
| 0.01s | 55.8 | 53.2 | 46.8 | 47.5 | 41.6 | 35.8 | 33.7 | 23.8 | 36.0 | 37.2 | E | 26.8 | 30.8 | E | E |
| 0.005s | 55.9 | 49.4 | 45.0 | 40.5 | 40.6 | 32.2 | 25.9 | 22.8 | 33.1 | 33.1 | E | 26.7 | E | E | E |
| 0.001s | 39.6 | 38.1 | 25.0 | 30.1 | 26.4 | 26.5 | E | E | E | E | E | E | E | E | E |

Error Detection

(b)

**Fig. 9:** **(a)** Chirp Template Matching. **(b)** Chirp Detection vs. Duration and Distance.

ChordMics has a good scalability because the limitation of wire connection disappears. It can be expected that ChordMics will deliver better performance if we deploy more microphones.

## D. *Chirp Detection Accuracy*

In this section, we evaluate the chirp detection accuracy of ChordMics. As the first step of ChordMics, chirp detection accuracy has a significant impact on the overall performance. In the experiment, the frequency range of the Chirp signal is 2KHz - 20KHz. We broadcast chirps at a predefined period of $T_p$. If we perfectly detect the chirps, the number of samples between two adjacent chirps will be $T_p \times F_s$ (see Figure 9(a)). We adopt mean absolute value (MAE) to evaluate the chirp detection, which is defined as $E(|\phi - T_p \times F_s|)$, where $\phi$ is the number of samples between the detected chirps.

Fig.9(b) summarizes how the chirp duration and distance between the chirp device and the microphone affect the chirp detection accuracy. In the figure, the number in each grid denotes the SNR under certain setting. The symbol 'E' means a serious error happened and the chirp detection fails. We find that as long as the chirp peak is larger than noise floor and noise spikes, the detection accuracy is satisfying. For all non-'E' grids, the MAE is always less than 1.5 samples (about 0.035ms). However, when SNR<25dB, the chirp detection tends to fail. In our design, we select 0.01s as the default duration of chirps because the coverage of it can be up to 30 m and, more importantly, it's almost inaudible and hardly impacts the recordings.

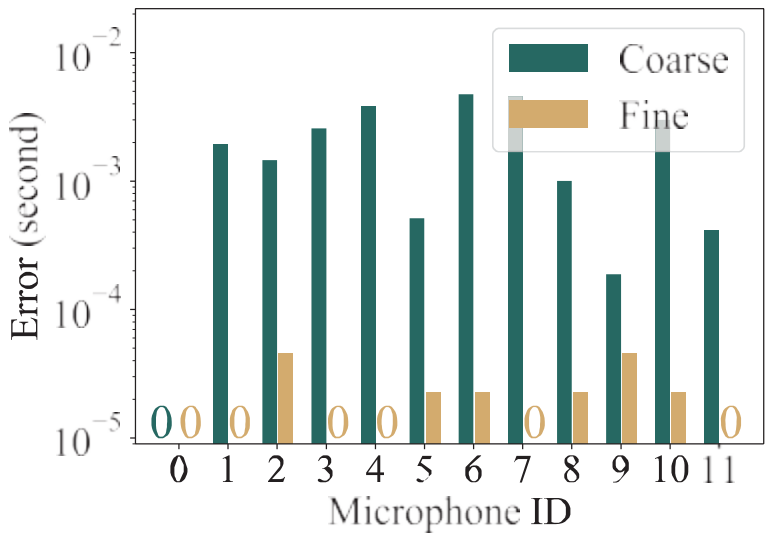

**Fig. 10:** Error of coarse-grained and fine-grained alignments

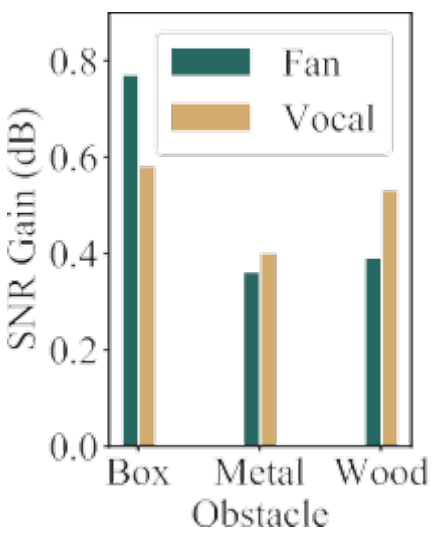

**Fig. 11:** Additional SNR gain for NLOS

## E. *Alignment Error*

In this section, we evaluate the accuracy of signal alignment, which is the key of ChordMics. In the experiment, to obtain the ground truth of relative delays, we equip each desired source with a chirp device which transmits chirp signal periodically. To avoid the interference between this chirp device and the reference device, the chirps transmitted by these two devices are set to different frequency bands.

Fig. 10 shows the estimation error of the relative delays after coarse-grained alignment and fine-grained alignment across 12 microphones. The 0-th microphone is considered as the reference. We can see that the error of coarse-grained alignment range from 8 samples to 210 samples. In other words, its error is less than 4.7ms (210/44100s). This means that the spatial resolution of our implementation is less than 1.5m, according to Eq. (10). The result also shows that the accuracy of the fine-grained alignment is two orders of magnitude smaller than that of the coarse-grained alignment. Near half of the microphones are perfectly aligned after fine-grained alignment, and the maximum deviation is just 2 samples (about 0.045ms).

## F. *Performance in NLOS scenario*

To alleviate multiple-path effect and mitigate NLOS problem, ChordMics tries to locate arriving signal paths corresponding to the desired source by identifying several desired correlation peaks, and constructively combines them to improve output signal strength. We run the experiment to evaluate the performance of ChordMics under NLOS scenarios, where three types of obstacles are placed between one microphone and the desired source. These obstacles include a paper box (1cm thickness), a metal plate (1.5cm thickness) and a wooden plank(1.2cm thickness). We compare the SNR gain of ChordMics (which combines the signal from multiple paths for signal enhancement) with the method which enhances the signal from a single path. Fig. 11 shows the SNR gains of two types of desired sources (industry fan and human voice) with different obstacles. We note that the SNR gains of the metal plate and the wooden plank are less than that of the paper box. The reason is that both the metal plate and the wooden plank cause serious attenuation, and the desired signal received by the NLOS microphone at these two cases is weaker. Despite the serious signal attenuation, ChordMics still improves performance by 0.36-0.77dB for single microphone.

## VI. Related Work

There have been many efforts made to enhance acoustic signal using microphone array and beamforming technique. Typical approaches include MBF [11], MVDR [10], [12], GSC [13], [14] and MWF [15]. However, all the above methods use centralized microphone arrays and thus provide limited coverage and low spatial resolution. Compared with these classical methods, ChordMics uses distributed microphone array, and thus provides higher spatial resolution, larger coverage, and better flexibility, without requiring accurate coordinate information and tight synchronization among microphones.

There are many distributed systems using collaborative nodes for beamforming or mitigrating interference. For example, MegaMIMO [16], [17] delivers a full-fledged PHY which supports distributed MIMO in real-time. PushID [18] exploits collaboration between graphically separated readers to enhance the range of RFID tags. These systems work well in the scenario where transmitters could get feedback from the receivers and (or) other transmitters for tuning the phase offsets. [19], [20] exploit interference pattern for concurrent transmission or scalable flooding. In our scenario, however, ChordMics cannot expect the transmitters (the desired acoustic sources), which are out of control, to cooperate with the microphones for distributed beamforming, such as sending preambles for channel estimation.

[21]–[24] propose distributed beamforming algorithms for acoustic enhancement. However, some practical issues (e.g. time offset) are not considered in those works. There are also some synchronization schemes dedicating in solving the time offset problem [25], [26]. However, most of those works remain on theoretical level with unrealistic assumptions (the co-variance of the noise is known as a prior), and are evaluated through simulation. They might be unavailable in practice.

## VII. Conclusion

We design a fully distributed microphone array to purify acoustic signals. Many fundamental challenges in distributed beamforming, like serious time offset among the microphones and the inaccurate coordinates of the microphones, are tackled by a coarse-to-fine signal alignment approach. We implement and evaluate ChordMics with extensive experiments, demonstrating the practicality and effectiveness of ChordMics.

## Acknowledgment

This work was supported by National Key R&D Program of China No. 2017YFB1003000, National Natural Science Foundation of China No. 61902213.

## References

[1] J. Guo, Y. He, and X. Zheng, "Pangu: Towards a software-defined architecture for multi-function wireless sensor networks," in *Proceedings of IEEE ICPADS*, 2017.

[2] Y. He, J. Guo, and X. Zheng, "From surveillance to digital twin: Challenges and recent advances of signal processing for industrial internet of things," *IEEE Signal Processing Magazine*, vol. 35, no. 5, pp. 120–129, 2018.

[3] "The acoustic signal recorded in a factory." [Online]. Available: https://www.dropbox.com/s/aghczzlefz55otu/sound.wav

[4] Z. Yu, C. Jiang, Y. He, X. Zheng, and X. Guo, "Crocs: Cross-technology clock synchronization for wifi and zigbee." in *Proceedings of EWSN*, 2018.

[5] Y. Wang, Y. Liu, Y. He, X.-Y. Li, and D. Cheng, "Disco: Improving packet delivery via deliberate synchronized constructive interference," *IEEE Transactions on Parallel and Distributed Systems*, vol. 26, no. 3, pp. 713–723, 2014.

[6] D. H. J. and D. D. E, *Array signal processing: concepts and techniques*. PTR Prentice Hall Englewood Cliffs, 1993.

[7] B. M. S and S. H. F, "A robust method for speech signal time-delay estimation in reverberant rooms," in *Proceedings of IEEE ICASSP*, 1997.

[8] L. E. A and M. D. G, *Digital communication*. Springer Science & Business Media, 2012.

[9] T. David and V. Pramod, *Fundamentals of wireless communication*. Cambridge university press, 2005.

[10] C. Jack, "High-resolution frequency-wavenumber spectrum analysis," *Proceedings of the IEEE*, vol. 57, no. 8, pp. 1408–1418, 1969.

[11] D. Rabinkin, R. Renomeron, J. Flanagan, and D. F. Macomber, "Optimal truncation time for matched filter array processing," in *Proceedings of IEEE ICASSP*, 1998.

[12] O. L. Frost, "An algorithm for linearly constrained adaptive array processing," *Proceedings of the IEEE*, vol. 60, no. 8, pp. 926–935, 1972.

[13] L. Griffiths and C. Jim, "An alternative approach to linearly constrained adaptive beamforming," *IEEE Transactions on antennas and propagation*, vol. 30, no. 1, pp. 27–34, 1982.

[14] S. Gannot, D. Burshtein, and E. Weinstein, "Signal enhancement using beamforming and nonstationarity with applications to speech," *IEEE Transactions on Signal Processing*, vol. 49, no. 8, pp. 1614–1626, 2001.

[15] S. Doclo and M. Moonen, "Gsvd-based optimal filtering for single and multimicrophone speech enhancement," *IEEE Transactions on signal processing*, vol. 50, no. 9, pp. 2230–2244, 2002.

[16] H. Rahul, S. S. Kumar, and D. Katabi, "Megamimo: Scaling wireless capacity with user demand," in *Proceedings of ACM SIGCOMM*, 2012.

[17] H. Ezzeldin, R. Hariharan, A. M. A, and K. Dina, "Real-time distributed mimo systems," in *Proceedings of ACM SIGCOMM*, 2016.

[18] J. Wang, J. Zhang, R. Saha, H. Jin, and S. Kumar, "Pushing the range limits of commercial passive rfids," in *Proceedings of USENIX NSDI*, 2019.

[19] M. Jin, Y. He, X. Zheng, D. Fang, D. Xu, T. Xing, and X. Chen, "Exploiting interference fingerprints for predictable wireless concurrency," *IEEE Transactions on Mobile Computing*, 2020.

[20] Y. Wang, Y. He, X. Mao, Y. Liu, and X.-Y. Li, "Exploiting constructive interference for scalable flooding in wireless networks," *IEEE/ACM Transactions on Networking*, vol. 21, no. 6, pp. 1880–1889, 2013.

[21] Y. Zeng and R. C. Hendriks, "Distributed delay and sum beamformer for speech enhancement via randomized gossip," *IEEE/ACM Transactions on Audio, Speech, and Language Processing*, vol. 22, no. 1, pp. 260–273, 2013.

[22] R. Heusdens, G. Zhang, R. C. Hendriks, Y. Zeng, and W. B. Kleijn, "Distributed mvdr beamforming for wireless microphone networks using message passing," in *International Workshop on Acoustic Signal Enhancement*, 2012.

[23] M. O'Connor and W. B. Kleijn, "Diffusion-based distributed mvdr beamformer," in *Proceedings of IEEE ICASSP*, 2014.

[24] M. Souden, K. Kinoshita, M. Delcroix, and T. Nakatani, "Location feature integration for clustering-based speech separation in distributed microphone arrays," *IEEE/ACM Transactions on Audio, Speech, and Language Processing*, vol. 22, no. 2, pp. 354–367, 2013.

[25] P. Pertilä, M. S. Hämäläinen, and M. Mieskolainen, "Passive temporal offset estimation of multichannel recordings of an ad-hoc microphone array," *IEEE/ACM Transactions on Audio, Speech, and Language Processing*, vol. 21, no. 11, pp. 2393–2402, 2013.

[26] S. Wehr, I. Kozintsev, R. Lienhart, and W. Kellermann, "Synchronization of acoustic sensors for distributed ad-hoc audio networks and its use for blind source separation," in *Proceedings of IEEE International Symposium on Multimedia Software Engineering*, 2004.